\documentclass[12pt]{article}
\usepackage{times}
\usepackage{fullpage}
\usepackage{amsmath}
\usepackage{amssymb}
\usepackage{amsfonts}
\usepackage{epsfig}
\usepackage[all,poly]{xy}

\newif\ifnotesw\noteswtrue
   {\ifnotesw\marginpar[\hfill\(\top\)]{\(\top\)}\fi}%
   {\ifnotesw\marginpar[\hfill\(\bot\)]{\(\bot\)}\fi}
   
      \newcommand{\mnote}[1]%
   {\ifnotesw\marginpar%
      [{\scriptsize\begin{minipage}[t]{\marginparwidth}
       \raggedleft#1%
       \end{minipage}}]%
       {\scriptsize\begin{minipage}[t]{\marginparwidth}
       \raggedright#1%
       \end{minipage}}%
    \fi}

\newcommand{\ignore}[1]{}

\newcommand{\bra}[1]{\langle #1 |}
\newcommand{\ket}[1]{| #1 \rangle}
\newcommand{\braket}[2]{\langle #1 | #2 \rangle}

\newcommand{\tderiv}[1]{\frac{\mathsf{d}}{\mathsf{d}t} #1}
\newcommand{\dt}{\mathsf{d}t}

\newcommand{\LL}{\mathbb{L}}
\newcommand{\UU}{\mathbb{U}}

\title{
Mixing and Decoherence in Continuous-Time Quantum Walks on Cycles
}
\author{
{Leonid Fedichkin}\footnote{Center for Quantum Device Technology, 
Department of Physics, 
and Department of Electrical and Computer Engineering,
Clarkson University, Potsdam, NY 13699--5721, USA. Email: leonid@clarkson.edu}
\and
{Dmitry Solenov}\footnote{Center for Quantum Device Technology
and Department of Physics, 
Clarkson University, Potsdam, NY 13699--5721, USA. Email: solenovd@clarkson.edu}
\and
{Christino Tamon}\footnote{Department of Mathematics and Computer Science 
and Center for Quantum Device Technology,
Clarkson University, Potsdam, NY 13699--5815, USA. Email: tino@clarkson.edu}
}
\date{\today}

\begin{document}
\thispagestyle{empty}
\bibliographystyle{plain}
\maketitle

\begin{abstract}
We prove analytical results showing that decoherence can be useful for mixing time in a continuous-time 
quantum walk on finite cycles. This complements the numerical observations by Kendon and Tregenna 
({\em Physical Review A} {\bf 67} (2003), 042315) of a similar phenomenon for discrete-time quantum walks.
Our analytical treatment of continuous-time quantum walks includes a continuous monitoring of all vertices 
that induces the decoherence process. We identify the dynamics of the probability distribution and observe 
how mixing times undergo the transition from quantum to classical behavior as our decoherence parameter grows 
from zero to infinity.
Our results show that, for small rates of decoherence, the mixing time improves linearly with decoherence, 
whereas for large rates of decoherence, the mixing time deteriorates linearly towards the classical limit.
In the middle region of decoherence rates, our numerical data confirms the existence of a unique optimal 
rate for which the mixing time is minimized.
\end{abstract}

\section{Introduction}

The study of quantum walks on graphs has gained considerable interest in quantum computation due to
its potential as an algorithmic technique and as a more natural physical model for computation.
As in the classical case, there are two important models of quantum walks, namely, the discrete-time
walks \cite{adz93, m96, aakv01, abnvw01}, and the continuous-time walks \cite{fg98,cfg02,ccdfgs03,cg03}.
Excellent surveys of both models of quantum walks are given in \cite{kendon,kempe}. 
In this work, our focus will be on continuous-time quantum walks on graphs and its dynamics under decoherence.

Some promising non-classical dynamics of continuous-time quantum walks were shown in \cite{mr02,k03,ccdfgs03}.
In \cite{mr02}, Moore and Russell proved that the continuous-time quantum walk on the $n$-cube achieves 
(instantaneous) uniform mixing in time $O(n)$, in contrast to the $\Omega(n \log n)$ time needed in the
classical random walk. Kempe \cite{k03} showed that the hitting time between two diametrically opposite
vertices on the $n$-cube is $n^{O(1)}$, as opposed to the well-known $\Omega(2^{n})$ classical bound
(related to the Ehrenfest urn model). In \cite{ccdfgs03}, an interesting algorithmic application of a
continuous-time quantum walk on a specific blackbox search problem was given. This latter result relied
on the exponentially fast hitting time of these quantum walks on path-collapsible graphs.

Further investigations on mixing times for continuous-time quantum walks were given in \cite{abtw03, gw03, aaht03}.
These works prove non-uniform (average) mixing properties for complete multipartite graphs, group-theoretic circulant
graphs, and the Cayley graph of the symmetric group. The latter graph was of considerable interest due to its potential 
connection to the Graph Isomorphism problem, although Gerhardt and Watrous's result in \cite{gw03} strongly discouraged
natural approaches based on quantum walks. All of these cited works have focused on unitary quantum walks, where
we have a closed quantum system without any interaction with its environment. 

A more realistic analysis of quantum walks that take into account the effects of decoherence was initiated by
Kendon and Tregenna \cite{kt03}.
In that work, Kendon and Tregenna made a striking numerical observation that a small amount of decoherence can be 
useful to improve the mixing time of discrete quantum walks on cycles. In this paper, we provide an analytical 
counterpart to Kendon and Tregenna's result for the continuous-time quantum walk on cycles. Thus showing that 
the Kendon-Tregenna phenomena is not merely an artifact of the discrete-time model, but suggests a fundamental 
property of decoherence in quantum walks. 
Recent realistic treatment for the hypercube was provided in a recent work by Alagi\'{c} and Russell \cite{ar05}.
Developing algorithmic applications that exploit this {\em positive} effect of decoherence on quantum mixing time 
provides an interesting challenge for future research.

In this work, we prove that Kendon and Tregenna's observation holds in the continuous-time quantum walk model.
Our analytical results show that decoherence can improve the mixing time in continuous-time quantum walk 
on cycles. We consider an analytical model due to Gurvitz \cite{g97} that incorporates the continuous monitoring 
of all vertices that induces the decoherence process. We identify the dynamics of probability distribution and 
observe how mixing times undergo transition from quantum to classical behavior as decoherence parameter grows 
from $0$ to $\infty$. For small rates of decoherence, we observe that mixing times improve linearly with decoherence, 
whereas for large rates, mixing times deteriorate linearly towards the classical limit. In the middle region of 
decoherence rates, we give numerical data that confirms the existence of a unique optimal rate for which the 
mixing time is minimal.


\section{Preliminaries}

Continuous-time quantum walks are well-studied in the physics literature 
(see, e.g., \cite{fls65}, Chapters 13 and 16), but mainly over constant-dimensional lattices.
It was studied recently by Farhi, Gutmann, and Childs \cite{fg98,cfg02} in the algorithmic context.
Let $G = (V,E)$ be an undirected graph with adjacency matrix $A_{G}$.
The Laplacian of $G$ is defined as $\mathcal{L} = A_{G}-D$, where $D$ is a diagonal matrix with
$D_{jj}$ is the degree of vertex $j$\footnote{We have $D = kI$, if $G$ is $k$-regular.}. 
If the time-dependent state of the quantum walk is $\ket{\psi(t)}$, 
then, by the Schr\"{o}dinger's equation, we have
\begin{equation}
i \hslash \tderiv{\ket{\psi(t)}} = \mathcal{L} \ket{\psi(t)}.
\end{equation}
The solution of the above equation is $\ket{\psi(t)} = e^{-it \mathcal{L}} \ket{\psi(0)}$ (assuming $\hslash = 1$).

We consider the $N$-vertex cycle graph $C_{N}$ whose adjacency matrix $A$ is a circulant matrix. 
The eigenvalues of $A$ are $\lambda_{j} = 2\cos(2\pi j/N)$ with corresponding eigenvectors $\ket{v_{j}}$, 
where $\braket{k}{v_{j}} = \frac{1}{\sqrt{N}}\exp(-2\pi i jk/N)$, for $j = 0,1,\ldots,N-1$. 
So, if the initial state of the quantum walk is $\ket{\psi(0)} = \ket{0}$, then $\ket{\psi(t)} = e^{-it L}\ket{0}$. 
After decomposing $\ket{0}$ in terms of the eigenvectors $\ket{v_{j}}$, we get
\begin{equation}
\ket{\psi(t)} 
        = e^{2it} \frac{1}{\sqrt{N}} \sum_{j=0}^{N-1} e^{-it\lambda_{j}}\ket{v_{j}}.
\end{equation}
The scalar term $e^{2it}$ is an irrelevant phase factor which can be ignored.

\begin{figure}[h]
\begin{center}
\epsfig{file=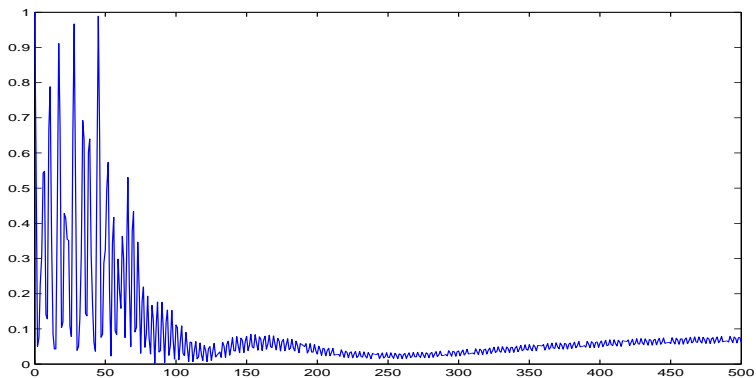, height=5cm, width=10cm}
\end{center}
\caption{Continuous-time quantum walk on the cycle $C_{5}$.
This is a plot of $|\psi_{0}(t)|^{2}$ for $t \in [0,500]$.
It exhibits a short-term chaotic behavior and a long-term oscillatory behavior.}
\label{figure:plot2}
\end{figure}

If $\ket{\psi(t)}$ represents the state of the particle at time $t$, 
let $P_{j}(t) = |\braket{j}{\psi(t)}|^{2}$ be the probability that the particle is at vertex $j$ at time $t$. 
Let $P(t)$ be the (instantaneous) probability distribution of the quantum walk on $G$.
The {\em average} probability of vertex $j$ over the time interval $[0,T]$ is defined as by
$\overline{P}_{j}(T) = \frac{1}{T} \int_{0}^{T} P_{j}(t) \ \dt$.
Let $\overline{P}(T)$ be the (average) probability distribution of the quantum walk on $G$ over the time
interval $[0,T]$.

To define the notion of mixing times of continuous-time quantum walks, we use the {\em total variation} 
distance between distributions $P$ and $Q$ that is defined as \ $||P - Q|| = \sum_{s} |P(s) - Q(s)|$.
For ~$\varepsilon \ge 0$, the ~$\varepsilon$-mixing time $T_{mix}(\varepsilon)$ of a continuous-time quantum 
walk is the minimum time $T$ so that ~$||P(T) - U_{G}|| \le \varepsilon$, where $U_{G}$ is the uniform distribution 
over $G$, or
\begin{equation}
T_{mix}(\varepsilon) \ = \ 
	\min\left\{T \ : \ \sum_{j=0}^{N-1} \left|P_{j}(T) - \frac{1}{N}\right| \le \varepsilon \right\}.
\end{equation}

\paragraph{Gurvitz's Model}
To analyze the decoherent continuous-time quantum walk on $C_{N}$, we use an analytical model developed by 
Gurvitz \cite{g97,gfmb03}. In this model, we consider the density matrix $\rho(t) = \ket{\psi(t)}\bra{\psi(t)}$
and study its evolution under a continuous monitoring of all vertices of $C_{n}$. Note that in this case, the
probability distribution $P(t)$ of the quantum walk is specified by the diagonal elements of $\rho(t)$, that is,
$P_{j}(t) = \rho_{j,j}(t)$.

The time-dependent non-unitary evolution of $\rho(t)$ in the Gurvitz model is given by (see \cite{smallG}):
\begin{equation}\label{drdt}
\tderiv{\rho_{j,k}(t)}  
	= i \ \left[\frac{\rho _{j,k+1} - \rho_{j+1,k} - \rho_{j-1,k} + \rho_{j,k-1}}{4}\right]
	- \Gamma \left({1 - \delta_{j,k}}\right)\rho_{j,k}
\end{equation}
Our subsequent analysis will focus on the variable $S_{j,k}$ defined as
\begin{equation}\label{6}
S_{j,k} = i^{k-j} \rho_{j,k}
\end{equation}
The above substitution reduces the system differential equations with complex coefficients into the following
system with only real coefficients:
\begin{equation}\label{dSdt}
\tderiv{S_{j,k}} = \frac{1}{4}\left({S_{j,k+1} + S_{j+1,k} - S_{j-1,k} - S_{j,k-1}}\right) -
	\Gamma \left({1 - \delta_{j,k}}\right) S_{j,k}.
\end{equation}
Throughout the rest of this paper, we will focus on analyzing Equation (\ref{dSdt}) for various rates of $\Gamma$.
One can note that, if $\Gamma=0$, there is an exact mapping of the quantum walk on a cycle onto a 
classical random walk on a two-dimensional torus. If $\Gamma \ne 0$, there is still an exact mapping of the 
quantum walk on a cycle onto some classical dynamics on a directed toric graph. 
This observation may be useful in estimating quantum speedup in other systems.

\begin{figure}[t]
\begin{center}
\epsfig{file=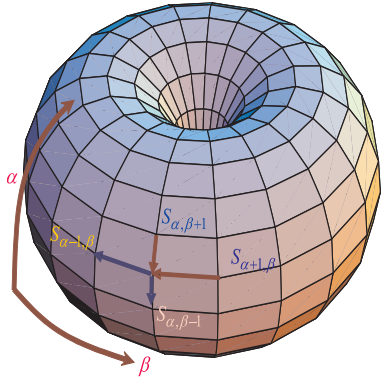, height=5cm, width=5cm}
\end{center}
\caption{The classical recurrence on the 2-dimensional torus derived from a decoherent continuous-time
quantum walk on the cycle.}
\label{figure:torus}
\end{figure}


\section{Small Decoherence}

We consider the decoherent continuous-time quantum walks when the decoherence rate $\Gamma$ is small.
More specifically, we consider the case when $\Gamma N \ll 1$.
First, we rewrite 
(\ref{dSdt}) as the perturbed linear operator equation
\begin{equation} \label{eqn:operator}
\tderiv{S(t)} = (\LL + \UU) \ S(t),
\end{equation}
where the linear operators $\LL$ and $\UU$ are defined as
\begin{eqnarray} \label{89}
\LL_{(\alpha,\beta)}^{(\mu,\nu)} 
	& = & 
	\frac{1}{4}\left( {\delta_{\alpha,\mu} \delta_{\beta,\nu - 1} + \delta_{\alpha,\mu - 1} \delta_{\beta,\nu} 
	- \delta_{\alpha,\mu} \delta_{\beta,\nu + 1} - \delta_{\alpha,\mu + 1} \delta_{\beta,\nu}} \right) \\
\UU_{(\alpha,\beta)}^{(\mu,\nu)} 
	& = &  
	- \Gamma \delta_{\alpha,\mu} \delta_{\beta,\nu} \left( {1 - \delta_{\alpha,\beta} } \right).
\end{eqnarray}
Here, we consider $\LL$ as a $N^2 \times N^2$ matrix where $\LL_{(\alpha,\beta)}^{(\mu,\nu)}$ 
is the entry of $\LL$ indexed by the row index $(\mu,\nu)$ and the column index $(\alpha,\beta)$. We view
$\UU$ in a similar manner.
The solution of (\ref{eqn:operator}) is given by $S(t) = e^{t(\LL + \UU)}S(0)$, ~or
\begin{equation} \label{7}
\tderiv{S_{\alpha,\beta}} = 
	\sum_{\mu,\nu = 0}^{N-1} \left({\LL_{(\alpha,\beta)}^{(\mu,\nu)} 
	+ \UU_{(\alpha,\beta)}^{(\mu,\nu)}} \right) S_{\mu,\nu},
\end{equation}
where $0 \le \alpha,\beta,\mu,\nu \le N-1$. 
The initial conditions are 
\begin{equation}\label{10}
\rho_{\alpha,\beta}(0) = S_{\alpha,\beta}(0) = \delta_{\alpha,0} \delta_{\beta,0}.
\end{equation}

\paragraph{Perturbation Theory}
We will use tools from the perturbation theory of linear operators (see \cite{kato,horn-johnson}).
To analyze Equation (\ref{eqn:operator}), we find the eigenvalues and eigenvectors of $\LL + \UU$.
Suppose that $V$ is some eigenvector of $\LL$ with eigenvalue $\lambda$, that is, $\LL V = \lambda V$.
Considering the perturbed eigenvalue equation
\begin{equation}
(\LL + \UU) (V + \tilde{V}) = (\lambda + \tilde{\lambda}) \ (V + \tilde{V}),
\end{equation}
we drop the second-order terms $\UU \tilde{V}$ and $\tilde{\lambda} \tilde{V}$ to obtain the first-order approximation
\begin{equation} \label{eqn:perturbed}
\UU \ V + \LL \ \tilde{V} \ = \ \tilde{\lambda} V \ + \ \lambda \tilde{V}.
\end{equation}
By taking the inner product of the above equation with $V^{\dagger}$, and since $\LL$ is Hermitian,
we see that the eigenvalue perturbation term $\tilde{\lambda}$ is defined as
\begin{equation}
\tilde{\lambda} \ = \ V^{\dagger} \UU V.
\end{equation}

Let $\mathcal{E}_{\lambda}$ be an eigenspace corresponding to the eigenvalue $\lambda$ and 
let ~$\{V_{k} : \ k \in I\}$ be a set of eigenvectors of ~$\LL$ that spans ~$\mathcal{E}_{\lambda}$.
Let $V = \sum_{k \in I} c_{k} V_{k}$ be a unit vector in $\mathcal{E}_{\lambda}$.
Using Equation (\ref{eqn:perturbed}), we have 
\begin{equation}
\sum_{k \in I} c_{k} \UU V_{k} = \tilde{\lambda} \sum_{k \in I} c_{k} V_{k},
\end{equation}
and after taking the inner product with $V_{j}^{\dagger}$, we get ~
$\sum_{k \in I} c_{k} V_{j}^{\dagger} \UU V_{k} = \tilde{\lambda} c_{j}$.
If the linear combination is uniform, that is $c_{j} = c$, for all $j$, then 
the eigenvalue perturbation $\tilde{\lambda}$ is simply given by
\begin{equation} \label{eqn:tilde-lambda}
\tilde{\lambda} \ = \ 
\sum_{k \in I} V_{j}^{\dagger} \UU V_{k}.
\end{equation}
In the case when $\mathcal{E}_{\lambda}$ is one-dimensional or the matrix $\UU$ is diagonal under
all similarity actions $V_{j}^{\dagger} \UU V_{k}$, for $j, k \in I$, 
the correction to the eigenvalues is given by the diagonal term $\tilde{\lambda} = V^{\dagger} \UU V$.
Otherwise, we need to solve the system described by $det(\UU_{\lambda} - \tilde{\lambda} I) = 0$.

To analyze the equation $S'(t) = (\LL + \UU)S(t)$, for which the solution is $S(t) = \exp[t(\LL + \UU)] S(0)$, 
we express $S(0)$ as a linear combination of the eigenvectors of $\LL + \UU$, say $\{V_{j} + \tilde{V}_{j}\}$.
In our case, the evolution of $S(t)$ can be described using the eigenvectors of $\LL$, since the contribution 
of the terms $\tilde{V}_{j}$ are negligible.
If ~$S(0) = \sum_{j} c_{j} V_{j}$, where $V_{j}$ are the eigenvectors of $\LL$, then
\begin{equation}
S(t) = \sum_{\lambda} e^{t(\lambda + \tilde{\lambda})} \sum_{j \in \mathcal{E}_{\lambda}} c_{j} \ V_{j}.
\end{equation}

\paragraph{Spectral Analysis}
The unperturbed linear operator $\LL$ has eigenvalues 
\begin{equation} \label{13}
\lambda_{(m,n)} 
	= i \ \sin\left(\frac{{\pi (m + n)}}{N}\right) \cos\left(\frac{{\pi (m - n)}}{N}\right)
\end{equation}
with corresponding eigenvectors
\begin{equation}\label{14}
V_{(\mu,\nu)}^{(m,n)} = \frac{1}{N} \exp\left(\frac{2\pi i}{N}(m\mu + n\nu)\right).
\end{equation}
Thus, for ~$0 \le m,n \le N-1$, we have
\begin{equation} \label{12}
\sum_{\mu,\nu = 0}^{N-1} {\LL_{(\alpha,\beta)}^{(\mu,\nu)} V_{(\mu,\nu)}^{(m,n)} }
	= \lambda_{(m,n)} V_{(\alpha,\beta)}^{(m,n)}.
\end{equation}
To analyze the effects of $\UU$, we compute the similarity actions of the eigenvectors on ~$\UU$:
\begin{eqnarray} 
\UU_{(m,n),(m',n')}
	& = & (V^{(m,n)})^{\dagger} \UU V^{(m',n')} \\
	& = & - \frac{\Gamma}{N^{2}} \sum_{(a,b)} (1 - \delta_{a,b}) \exp\left(\frac{2\pi i}{N}[(m' - m)a + (n' - n)b]\right) \\
\label{eqn:U-matrix}
	& = & - \Gamma \ \delta_{m',m} \ \delta_{n',n} 
		+ \frac{\Gamma}{N} \ \delta_{[(m'-m)+(n'-n)] ~(\mbox{\scriptsize mod $N$}), 0}
\end{eqnarray}
where $0 \le m,m',n,n' \le N-1$.\\

The eigenvalues $\lambda_{(m,n)}$ of $\LL$ have the following important {\em degeneracies}:
\begin{enumerate}
\item[(a)] Diagonal ($m = n$): $\lambda_{(m,m)} = i \ \sin(2\pi m/N)$. \\
	Each of this eigenvalue has multiplicity $2$, by the symmetries of the sine function.
	This degeneracy is absent in our case, since $\UU$ is diagonal over the corresponding eigenvectors. 
	For example, $\UU_{(m,n),(N/2-m,N/2-m)} = 0$, ~for ~$0 < m < N/2$.

\item[(b)] Zero ($m + n \equiv 0\pmod{N}$): $\lambda_{(m,n)} = 0$. \\
	This degeneracy is absent in our case since the corresponding eigenvectors are not involved in the
	linear combination of the initial state $S(0)$.

\item[(c)] Off-diagonal ($m \neq n$): $\lambda_{(m,n)} = \lambda_{(n,m)}$. \\
	Since $\lambda_{(m,n)} = i \ [\sin(2\pi m/N) + \sin(2\pi m/N)]$, each of this eigenvalue has multiplicity 
	at least 4, due to the symmetries of the sine function. 
	In our case, the effective degeneracy of these eigenvalues are 2, again by a similar argument.

	By (\ref{eqn:U-matrix}), the off-diagonal contribution is present if 
	~$m + n \equiv m' + n'\pmod{N}$. 
	Thus, ~$\lambda_{(m,n)} = \lambda_{(m',n')}$ ~implies that ~$\cos(\pi(m-n)/N) = \pm \ \cos(\pi(m'-n')/N)$,
	since ~$\sin(\pi(m+n)/N) = \pm \sin(\pi(m'+n')/N)$.
	This implies that ~$m-n = -(m'-n')$ or $|(m-n)-(m'-n')| = N$, since ~$-(N-1) \le m-n, m'-n' \le N-1$.
	In either case, we get $m = n \pm N/2$ or $m' = n' \pm N/2$.
	But, upon inspection, we note that $\UU$ is diagonal over these combinations, except for the case
	when $(m',n') = (n,m)$.
\end{enumerate}

In what follows, we calculate the eigenvalue perturbation terms $\tilde{\lambda}$.
For {\em simple} eigenvalues, these correction terms are given by the diagonal elements
\begin{equation}
\tilde{\lambda}_{(m,n)}
	\ = \ (V^{(m,n)})^{\dagger} \ \UU \ V^{(m,n)} 
	\ = \ -\Gamma \frac{(N-1)}{N},
\end{equation}
by Equation (\ref{eqn:U-matrix}). 
For a {\em degenerate} eigenvalue $\lambda_{(m,n)}$ with multiplicity two, 
~if ~$V = c (V^{(m,n)} + V^{(n,m)})$, for some constant ~$c$,
~then ~$\tilde{\lambda}_{(m,n)} = (V^{(m,n)})^{\dagger} \UU V$,
and similarly for $V^{(n,m)}$.
Further calculations reveal that
the eigenvalue perturbation $\tilde{\lambda}_{(m,n)}$ is 
\begin{equation}
\tilde{\lambda}_{(m,n)} 
	\ = \ (V^{(m,n)})^{\dagger} \ \UU \ V^{(m,n)} + (V^{(m,n)})^{\dagger} \ \UU \ V^{(n,m)} 
	\ = \ -\Gamma \frac{(N-2)}{N},
\end{equation}
again by Equation (\ref{eqn:U-matrix}).

\paragraph{Dynamics}
We are ready to describe the full solution to Equation (\ref{dSdt}). First, note that there exists a
{\em trivial} time-independent solution given by $S^{0}_{\alpha,\beta}(t) = \frac{\delta_{\alpha,\beta}}{N}$,
that can be expressed as the following linear combination of the eigenvectors of $\LL$:
\begin{equation}
S^{0}(t) = \sum_{(m,n)} \frac{1}{N} \ (\delta_{m+n,0} + \delta_{m+n,N}) \ V^{(m,n)}.
\end{equation}
The particular solution will depend on the initial condition $S(0)$, 
where $S_{\alpha,\beta}(0) = \delta_{\alpha,0}\delta_{\beta,0}$.
Note that we have
\begin{equation}
S(0) = \sum_{(m,n)} \frac{1}{N} \ V^{(m,n)}.
\end{equation}
Thus, the solution is of the form
\ignore{
\begin{equation}
S(t) = S^{0}(t) + 
	\frac{1}{N} \sum_{(m,n)} (1 - \delta_{[m+n] (\mbox{\scriptsize mod $N$}),0}) \ 
	e^{t(\lambda_{(m,n)} + \tilde{\lambda}_{(m,n)})} \ V^{(m,n)}.
\end{equation}
This implies that
}
\begin{equation}
S_{\alpha,\beta}(t) 
	= \frac{\delta_{\alpha,\beta}}{N} + 
	\frac{1}{N^{2}} \sum_{(m,n)} 
	(1 - \delta_{[m+n] (\mbox{\scriptsize mod }N),0}) \ 
	e^{t(\lambda_{(m,n)} + \tilde{\lambda}_{(m,n)})} \ 
	\exp\left[\frac{2\pi i}{N}(m\alpha + n\beta)\right]
\end{equation}
The probability distribution of the continuous-time quantum walk is given by the diagonal terms
$P_{j}(t) = S_{j,j}(t)$, that is
\begin{eqnarray*}
P_{j}(t)
	& = & \frac{1}{N} + 
		\frac{1}{N^{2}} \sum_{(m,n)} (1 - \delta_{m+n(\mbox{\scriptsize mod }N),0}) 
		\times \left[\delta_{m,n} e^{-\Gamma \frac{N-1}{N}t} + (1-\delta_{m,n}) e^{-\Gamma \frac{N-2}{N}t}\right] \\
	&   & \times \exp\left[i\sin\left(\frac{\pi(m+n)}{N}\right)\cos\left(\frac{\pi(m-n)}{N}\right)\right] 
		\exp\left[\frac{2\pi i}{N}(m + n)j\right] 
\end{eqnarray*}
We calculate an upper bound on the $\varepsilon$-uniform mixing time $T_{mix}(\varepsilon)$.
For this, we define 
\begin{equation}
M_{j}(t) = \frac{1}{N} \sum_{m=0}^{N-1} e^{it\sin(2\pi m/N)} \omega_{N}^{mj},
\end{equation}
where $\omega_{N} = \exp(2\pi i/N)$.
Note that 
\begin{equation}
M_{j}^{2}(t/2) = \frac{1}{N^{2}} \sum_{m,n=0}^{N-1} e^{it\lambda_{(m,n)}} \omega_{N}^{(m+n)j}, \ \ \
M_{2j}(t) = \frac{1}{N} \sum_{m=0}^{N-1} e^{it\lambda_{(m,m)}} \omega_{N}^{2mj}
\end{equation}
Using these expressions, 
we have
\begin{eqnarray}
\left|P_{j}(t) - \frac{1}{N}\right| 
	& \le & e^{-\Gamma \frac{N-2}{N}t} \left| M_{j}^{2}(t/2) + \frac{e^{-t\Gamma/N} - 1}{N}
		\left[M_{2j}(t) - \frac{2 - (N \mbox{ mod } 2)}{N}\right] \right| \\
	& \le & e^{-\Gamma \frac{N-2}{N}t} \ \left|1 + \frac{e^{-t\Gamma/N} - 1}{N} (1 - 2/N)\right|.
\end{eqnarray}
One can note that $|M_{j}(t)| \le 1$, and therefore,
\begin{equation}
\sum_{j=0}^{N-1} \left|P_{j}(t) - \frac{1}{N}\right| 
	\ \le \
	e^{-\Gamma \frac{N-2}{N}t} \ (N + e^{-t\Gamma/N} - 1).
\end{equation}
Since $e^{-t\Gamma/N} \le 1$, the above equation shows that
$N e^{-\Gamma \frac{N-2}{N}t} \le \varepsilon$. This gives the mixing time bound of
\begin{equation}
T_{mix}(\varepsilon) 
	\ < \ 
	\frac{1}{\Gamma} 
	\ln\left(\frac{N}{\varepsilon}\right)
	\left[1 + \frac{2}{N-2}\right].
\end{equation}


\section{Large Decoherence}

We analyze the decoherent continuous-time quantum walks when the decoherence rate $\Gamma$ is large,
that is, when $\Gamma \gg 1$.
In our analysis, we will focus on diagonal sums of the matrix $S(t)$ from (\ref{dSdt}). 
For $k = 0,\ldots,N-1$, we define the diagonal sum $D_{k}$ as
\begin{equation}
D_{k}  = \sum_{j=0}^{N-1} S_{j, ~j+k \mbox{\scriptsize ~(mod $N$)}},
\end{equation}
where the indices are treated as integers modulo $N$. We note that
\begin{equation}\label{dDdt}
\tderiv{D_{k}} = - \Gamma \left( {1 - \delta _{k,0}} \right) D_{k}.
\end{equation}
We refer to the diagonal $D_{0}$ as {\em major} and the other diagonals as {\em minor}.
Equation (\ref{dDdt}) suggests that the minor diagonal sums decay strongly with characteristic time 
of order ~$1/\Gamma$. By the initial conditions, the non-zero elements appear only along the major diagonal.
From (\ref{dSdt}), it follows that the system will evolve initially in the following way.
The elements on the two minor diagonals {\em nearest} to the major diagonal will deviate slightly away from 
zero due to nonconformity of classical probability distribution along the major diagonal. 
This process with a rate of order ~$1/4$ will compete with a self-decay with rate of order ~$\Gamma \gg 1/4$,
thereby limiting the corresponding off-diagonal elements to small values of the order ~$1/\Gamma$.
A similar argument applies to elements on the other minor diagonals which will be kept very small compared 
to their neighbors that are closer to the major diagonal and will be of the order of ~$1/\Gamma^2$, etc.
By retaining only matrix elements that are of order of ~$1/\Gamma$, we derive a truncated set of 
differential equations for the elements along the major and the two adjacent minor diagonals:
\begin{eqnarray}
\label{dSaa}
{S'_{j,j}}   & = & \frac{1}{4}\left(S_{j,j+1} + S_{j+1,j} - S_{j-1,j} - S_{j,j-1}\right), \\
\label{dSaa1}
{S'_{j,j+1}} & = & \frac{1}{4}\left(S_{j+1,j+1} - S_{j,j}\right) - \Gamma S_{j,j+1}, \\
\label{dSa1a}
{S'_{j,j-1}} & = & \frac{1}{4}\left(S_{j,j} - S_{j-1,j-1}\right) - \Gamma S_{j,j-1}.
\end{eqnarray}
To facilitate our subsequent analysis, we define
\begin{equation}
a_j = S_{j,j}, \ \ \ \ \ d_j = S_{j,j+1} + S_{j+1,j}.
\end{equation}
Then, we observe that
\begin{equation}
a'_{j} = \frac{\left(d_j - d_{j-1} \right)}{4}, \ \ \ \ \
d'_{j} = \frac{\left(a_{j+1} - a_j\right)}{2} - \Gamma d_j.
\end{equation}
The general solution of the above system of difference equations has the form
\begin{eqnarray}
a_j & = & \frac{1}{N}\sum_{k=0}^{N-1} \
	\left\{A_{k,1}\exp{\left(-\gamma_{k,0}t\right)} + A_{k,2}\exp{\left(-\gamma_{k,1}t\right)}\right\} \ \omega^{jk} \\
d_j & = & \frac{1}{N}\sum_{k=0}^{N-1} \
	\left\{D_{k,1}\exp{\left(-\gamma_{k,0}t\right)} + D_{k,2}\exp{\left(-\gamma_{k,1}t\right)}\right\} \ \omega^{jk}
\end{eqnarray}
where $\omega = e^{2\pi i/N}$, and the exponents $\gamma_{k,0}$ and $\gamma_{k,1}$ are the quadratic roots of
\begin{equation}
x(\Gamma-x) = \frac{1}{2}\sin^2\left({\frac{\pi k}{N}}\right).
\end{equation}
Letting $\gamma_{k,0} < \gamma_{k,1}$, we have
\begin{eqnarray}
\gamma_{k,0} & = & \frac{1}{2\Gamma}\sin^2\left(\frac{\pi k}{N}\right) + o\left(\frac{1}{\Gamma}\right), \\
\gamma_{k,1} & = & \Gamma - \frac{1}{2\Gamma}\sin^2\left(\frac{\pi k}{N}\right) + o\left(\frac{1}{\Gamma}\right).
\end{eqnarray}
By the initial conditions ~$a_j(0) = \delta_{j,0}$ ~and ~$d_{j}(0) = 0$, for $j = 0,\ldots,N-1$. Thus,
\ignore{
the equations:
\begin{equation}
A_{k,1} = 1 - A_{k,0}, \ \ \ D_{k,1} = - D_{k,0}
\end{equation}
\begin{equation}
A_{k,0} = \frac{\gamma_{k,1}}{\gamma_{k,1} - \gamma_{k,0}},  \ \ \
D_{k,0} = i\sin{\frac{\pi k}{N}} \exp{\left(\frac{i \pi k}{N}\right)} \frac{A_{k,1}}{\left(\Gamma - \gamma_{k,1}\right)}.
\end{equation}
By combining these equations, we have
}
\begin{equation}
A_{k,0} \ \simeq \ 1, \ \ \
A_{k,1} \ \simeq \ - \ \frac{1}{\Gamma^2} \sin^2{\frac{\pi k}{N}}
\end{equation}
and, for $b = 0,1$, we have
\begin{equation}
D_{k,b} \ \simeq \ (-1)^{b} \ \frac{i}{\Gamma} \sin\left(\frac{\pi k}{N}\right) \exp{\left(\frac{i \pi k}{N}\right)},
\end{equation}
These equations show that the amplitudes of the elements along minor diagonals are reduced by 
an extra factor of $\Gamma$ compared to the elements along the major diagonal. 
Summarizing, the solution of differential equation at large $\Gamma$ has the form
\begin{equation}
a_j = \frac{1}{N}\sum_{k=0}^{N-1} \
	\exp{\left(-\frac{\sin^2{\frac{\pi k}{N}}}{2\Gamma} t\right)} \ \omega^{jk}.
\end{equation}
Based on the above analysis, the full solution for $S(t)$ is given by
\begin{equation} \label{eqn:large-solution}
S_{j,k}(t) =
	\left\{\begin{array}{ll}
	a_{j}		& \mbox{ if ~$j = k$ } \\
	d_{j}/2		& \mbox{ if ~$|j - k| = 1$ } \\
	0		& \mbox{ otherwise }
	\end{array}\right.
\end{equation}
It can be verified that $S(t)$ is a solution to Equation (\ref{dSdt}) modulo terms of order ~$o(1/\Gamma)$.

The total variation distance between the uniform distribution and the probability distribution of the
decoherent quantum walk on $C_{N}$ is given by
\begin{equation} 
\sum_{j=0}^{N-1} \left| {a_j(t) - \frac{1}{N}} \right| = 
	\sum_{j=0}^{N-1} \left|\frac{1}{N} \sum_{k=0}^{N-1} 
		\exp\left(-\frac{\sin^2{\frac{\pi k}{N}}}{2\Gamma} t\right)
		\exp\left(\frac{2\pi ijk}{N}\right) - \frac{1}{N}\right|,
\end{equation}
which simplifies to
\begin{equation} 
\sum_{j=0}^{N-1} \left| {a_j(t) - \frac{1}{N}} \right| =
      \frac{1}{N} \sum_{j=0}^{N-1} \left| \sum_{k=1}^{N-1}
	\exp\left(-\frac{\sin^2{\frac{\pi k}{N}}}{2\Gamma} t\right)
	\cos{\left(\frac{2\pi k j}{N}\right)} \right|.
\end{equation}

\paragraph{Lower bound}
A lower bound on the mixing time for large decoherence rate $\Gamma$ can be derived as follows.
Note that
\begin{eqnarray}
\sum_{j=0}^{N-1} {\left| {a_j(t) - \frac{1}{N}} \right|}
	& \ge & {\left| {a_0(t) - \frac{1}{N}} \right|}
	=   \frac{1}{N} \sum_{k=1}^{N-1} \exp{\left(-\frac{\sin^2{\frac{\pi k}{N}}}{2\Gamma} t\right)}, \\
	& \ge & \frac{2}{N} \exp{\left(-\frac{\sin^2{\frac{\pi}{N}}}{2\Gamma} t\right)},
\end{eqnarray}
where the first inequality uses the term $j = 0$ only and the second inequality uses the terms $k = 1,N-1$.
This expression is monotone in $t$, and is a lower bound on the total variation distance.
It reaches $\varepsilon$ at time $T_{lower}$, when
\begin{equation}
T_{lower} 
	\ = \ 
	\frac{2\Gamma}{\sin^{2}{\frac{\pi}{N}}} \ln\left( \frac{2}{N \varepsilon} \right)
	\ \simeq \ 
	\frac{2\Gamma N^2}{\pi^2} \ln\left(\frac{2}{N\varepsilon}\right),
\end{equation}
for large $N \gg 1$.

\paragraph{Upper bound}
An upper bound on the mixing time for large decoherence rate $\Gamma$ can be derived as follows.
Consider the following derivation:
\begin{eqnarray} 
\sum_{j=0}^{N-1} {\left| {a_j(t) - \frac{1}{N}} \right|}
	& = & \frac{1}{N} \sum_{j=0}^{N-1} 
		{\left| {\sum_{k=1}^{N-1} \exp{\left(-\frac{\sin^2{\frac{\pi k}{N}}}{2\Gamma} t\right)} 
		\cos{\left(\frac{2\pi k j}{N}\right)}} \right|} \\
	& \le & \frac{1}{N} \sum_{j=0}^{N-1} 
		\sum_{k=1}^{N-1} \exp{\left(-\frac{\sin^2{\frac{\pi k}{N}}}{2\Gamma} t\right)},
\end{eqnarray} 
since $|\cos(x)| \le 1$. The last expression is equal to
\begin{eqnarray} 
\sum_{k=1}^{N-1} \exp{\left(-\frac{\sin^2{\frac{\pi k}{N}}}{2\Gamma} t\right)} 
	& =   & 2 \sum_{k=1}^{\lfloor N/2 \rfloor} 
		\exp{\left(-\frac{\sin^2{\frac{\pi k}{N}}}{2\Gamma} t\right)}, \\
	& \le & 2 \sum_{k=1}^{\lfloor N/2 \rfloor} 
		\exp{\left(-\frac{2 k^2 t}{\Gamma{N^2}} \right)},
\end{eqnarray} 
where the last inequality is due to ~$\sin(x) > 2x/\pi$, ~whenever ~$0 < x < \pi/2$ ~(see Eq. 4.3.79, \cite{as}).
Since $k \ge 1$, we have $k^2 \ge k$. Thus, we have
\begin{equation}
\sum_{j=0}^{N-1} {\left| {a_j(t) - \frac{1}{N}} \right|} 
	\ < \ 2 \sum_{k=1}^{\lfloor N/2 \rfloor} \exp{\left(-\frac{2 k t}{\Gamma{N^2}} \right)} 
	\ < \ 2 \sum_{k=1}^{\infty} \exp{\left(-\frac{2 kt}{\Gamma{N^2}} \right)}.
\end{equation}
The last expression is a geometric series that equals ~
$2/[\exp(2t/(\Gamma{N^2})) - 1]$.
This expression is monotone in $t$, and it is the upper bound for the total variation distance.
It reaches $\varepsilon$ value at time $T_{upper}$, when
\begin{equation}
T_{upper} \ = \ \frac{\Gamma N^2}{2} \ln\left(\frac{2+\varepsilon}{\varepsilon}\right).
\end{equation}


\section{Conclusions}

In this work, we studied the average mixing times in a continuous-time quantum walk on the $N$-vertex 
cycle $C_{N}$ under decoherence. For this, we used an analytical model developed by S. Gurvitz \cite{g97}.
We found two distinct dynamics of the quantum walk based on the rates of the decoherence parameter.
For small decoherence rates, where $\Gamma N \ll 1$, the mixing time is bounded as
\begin{equation}
T_{mix} \ < \ \frac{1}{\Gamma} \ln\left(\frac{N}{\varepsilon}\right) \left[1 + \frac{2}{N-2}\right]. 
\end{equation}
This bound shows that $T_{mix}$ is inversely proportional to the decoherence rate $\Gamma$. 
For large decoherence rates $\Gamma \gg 1$, the mixing times are bounded as 
\begin{equation}
\frac{\Gamma N^2}{\pi^2} \ln\left(\frac{2}{N \varepsilon}\right) 
\ < \ T_{mix} \ < \
\frac{\Gamma N^2}{2} \ln \left(\frac{2+\varepsilon}{\varepsilon}\right).
\end{equation}
These bounds are show that $T_{mix}$ is linearly proportional to the decoherence rate $\Gamma$,
but is quadratically dependent on $N$. Note that the dependences on $N$ of the mixing times
exhibit the expected quantum to classical transition.

\begin{figure}[h]
\begin{center}
\epsfig{file=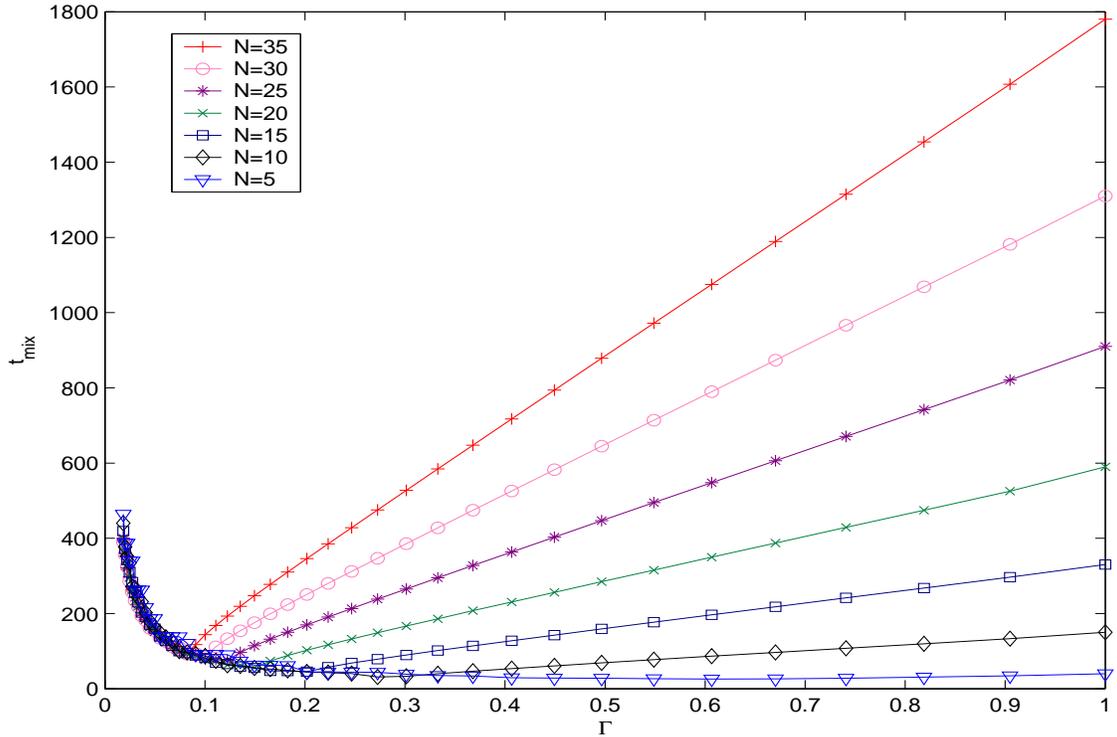, height=10cm, width=15cm}
\end{center}
\caption{
The quantum to classical transition of mixing time in a continuous-time decoherent quantum walk on $C_{N}$, 
for $N=5,10,15,20,25,30,35$.
}
\label{figure:dima}
\end{figure}

These analytical results already point to the existence of an {\em optimal} decoherence rate for which 
the mixing time is minimum. Our additional numerical experiments (see Figure (\ref{figure:dima})) 
for $\Gamma \sim 1$ confirmed that there is a unique optimal decoherence rate for which the mixing time 
is minimum. This provides a continuous-time analogue of the Kendon and Tregenna results in \cite{kt03}.

\section*{Acknowledgments}

We thank Viv Kendon for her kind encouragements in our interests on decoherence in continuous-time quantum 
walks and Vladimir Privman for helpful discussion. This research was supported by the National Science 
Foundation grant DMR-0121146.

\end{document}